\documentstyle[sprocl]{article}

\input{psfig}

\input mymacro.sty

\begin{document}

\title{
An Almost Perfect 
Lattice Action for 
infrared QCD 
\footnote{This talk is based on our recent works
~\cite{fuji00a,maxim00a,yama00a,shoji99,lat99}}
}

\author{T.Suzuki}

\address{
Institute for Theoretical Physics, Kanazawa University,\\
Kanazawa 920-1192, Japan\\
E-mail: suzuki@hep.s.kanazawa-u.ac.jp
} 

\maketitle

\abstracts{ A block-spin transformation 
on the dual lattice leads us to an
almost perfect lattice action for monopoles and strings in QCD.
The perfect operator for a static quark potential is fixed 
when we compare the above action with the perfect action obtained 
analytically after 
infinite-step block-spin transformations in a simple case.
The continuum rotational invariance is restored and the physical value
of the string tension is reproduced fairly well.
Gauge independence of the abelian and the monopole scenario is
discussed.
}

\section{Introduction}
Low-energy effective theory of QCD is important for analytical
understanding of hadron physics. 
One of approaches 
is to search for relevant dynamical variables and
to construct an effective theory for these variables.

From this point of view, the idea proposed by 't Hooft
\cite{'thooft} is very promising. The
confinement of quarks can be explained as the dual Meissner effect
which is due to condensation of these monopoles after 
an abelian projection.

Many numerical results
support the dual superconductor picture of confinement \cite{domi} in the
Maximal abelian (MA) gauge~\cite{Kronfeld} in
the framework of lattice QCD. We
expect hence, after integrating out all degrees of freedom other than
the monopoles, an effective theory described by the monopoles works
well in the IR region of QCD.

The effective monopole action on the MA projection of $SU(2)$ lattice
QCD was first obtained by Shiba and Suzuki \cite{shiba_suzuki}
using an inverse Monte-Carlo method ~\cite{Swendsen}.
See also Ref.\cite{nakam}. 

 The purpose  of this talk is to review briefly the results in our recent 
publications \cite{fuji00a,maxim00a,yama00a,shoji99,lat99}.

\section{An (almost) perfect monopole action}\label{PerfecAc}

The method to derive the monopole action is the following.
\begin{itemize}
\item[1] 
We generate link fields $\mbra{U(s,\mu)}$ using the
simple Wilson action for SU(2) and SU(3) QCD.
We consider $24^4$ and $48^4$ hyper-cubic lattice for $\beta=2.0\sim 2.8$
(SU(2)) and for $\beta=5.6\sim 6.4$ (SU(3)).
\item[2] 
Next we perform the abelian projection in the MA
gauge to separate abelian link variables $\{u(s,\mu)\}$.
\item[3] 
Monopole currents $k_\mu(s)$ can be defined from abelian link variables
following DeGrand and Toussaint\cite{D_T}.
\item[4]
We determine the set of couplings $G_i$ from the monopole current 
ensemble $\mbra{k_{\mu}(s)}$ with the aid of an inverse 
Monte-Carlo method first developed by Swendsen\cite{Swendsen} 
and extended to closed monopole currents by Shiba and Suzuki 
~\cite{shiba_suzuki}.
Here the monopole action can be written as a linear combination of some
operators
$ {\cal S}[k] = \sum_{i} G_i {\cal S}_i[k].$

Practically, we have to restrict the number of interaction terms. 
The form of actions adopted here
is 27 quadratic interactions and 4-point and 6-point interactions.
\item[5] 
We perform a block-spin transformation in terms of 
the monopole currents on the dual lattice 
to investigate the renormalization flow in the IR region. 
We adopt $n=1,2,3,4,6,8$ extended monopole currents
as an $n$ blocked operator\cite{ivanenko}:
\begin{eqnarray}
K_\mu(s^{(n)})&=&
  \sum_{i,j,l=0}^{n-1}
    k_\mu(ns^{(n)}+(n-1)\hat{\mu}+i\hat{\nu}+j\hat{\rho}+l\hat{\sigma})
\label{pfac:2}
\end{eqnarray}
The renormalized lattice spacing is $b=na(\beta)$ and the continuum limit is 
taken as the limit $n \to \infty$ for a fixed physical length $b$.

\item[6]
The physical length $b=n a(\beta)$ is taken in unit of the physical 
string tension $\sqrt{\sigma_{phys}}$. 
\end{itemize}

\section{Numerical results}\label{resul}
\begin{itemize}
\item[1.]
 The couplings are fixed 
clearly. We see  the scaling ${\cal S}[k_{\mu}, n,a(\beta)] \to
 {\cal S}[k_{\mu},b=na(\beta)]$
for fixed $b=na(\beta)$ looks 
good. The continuum limit is taken as 
$a\to 0,\ \ n\to \infty$ for $ b=n\cdot a$ fixed.
\item[2.] 
The four- and six-point interactions become negligibly 
small for IR region. 
Two-point interactions are relatively dominant
for large $b$ region.
\item[3.] 
We see the direction 
dependence of the current action from 
the data. 
For example, two nearest-neighbor interactions $G_2$ and $G_3$ 
are quite different for small $b$ region.

\begin{minipage}[tbh]{11cm}
\begin{flushleft}
  \epsfxsize=50mm
  \epsfysize=40mm
  \leavevmode
  \epsfbox{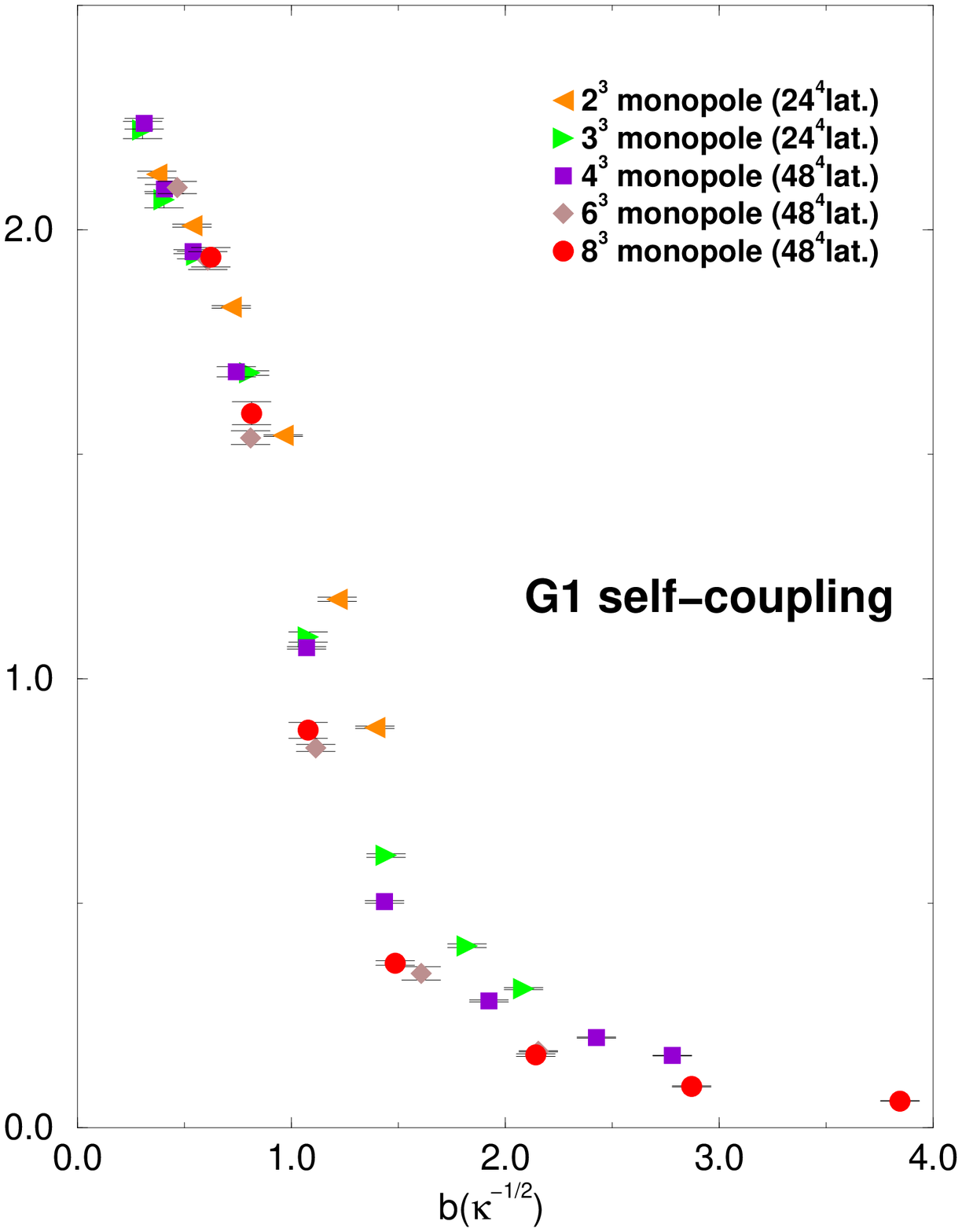}
\end{flushleft}

\vspace{-48mm}

\begin{flushright}
  \epsfxsize=50mm
  \epsfysize=40mm
  \leavevmode
  \epsfbox{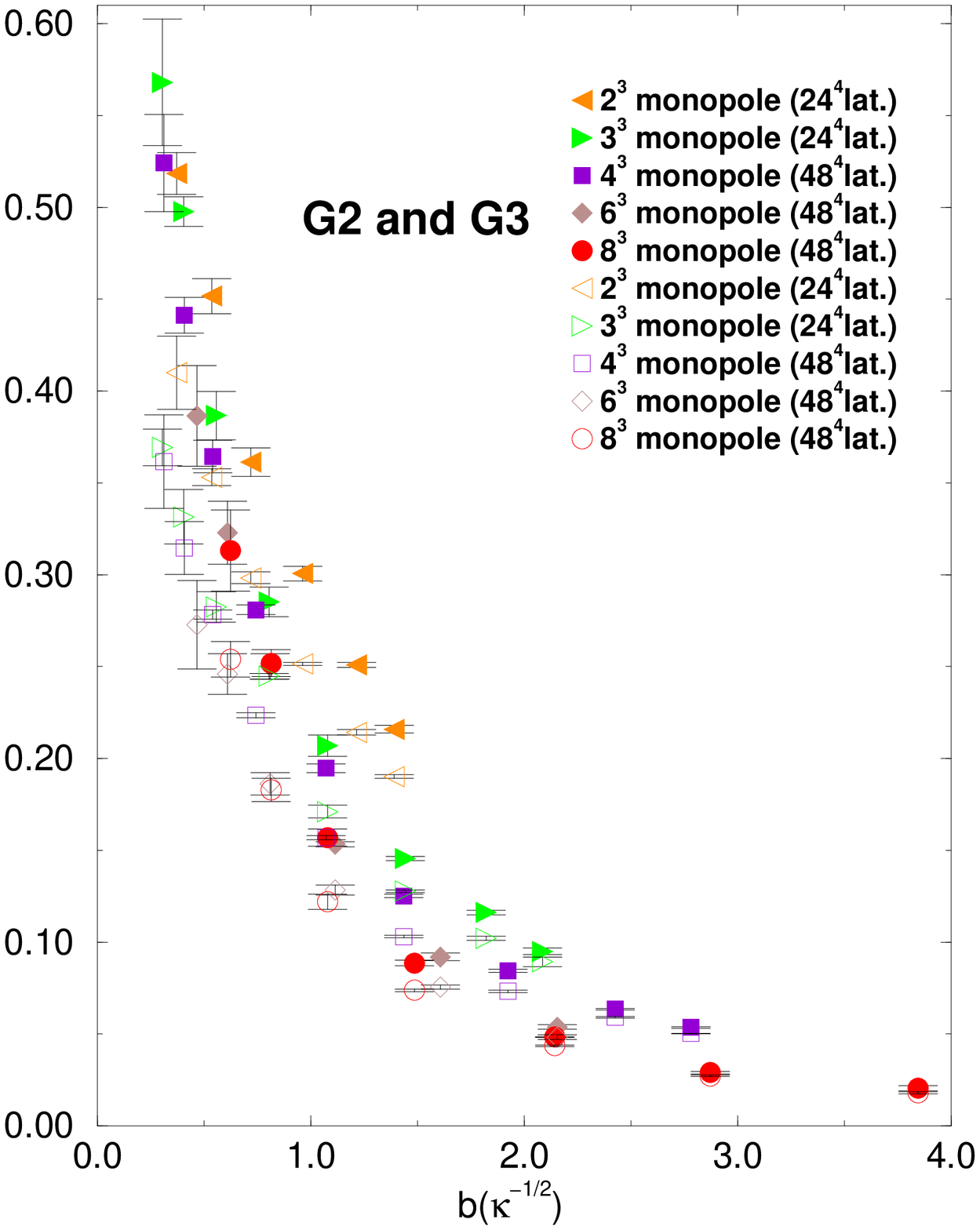}
\end{flushright}

Figure 1: The dominant couplings of quadratic interactions 
versus physical length $b$ in SU(2).

\end{minipage}

\begin{minipage}[tbh]{11cm}
\begin{flushleft}
  \epsfxsize=50mm
  \epsfysize=50mm
  \leavevmode
  \epsfbox{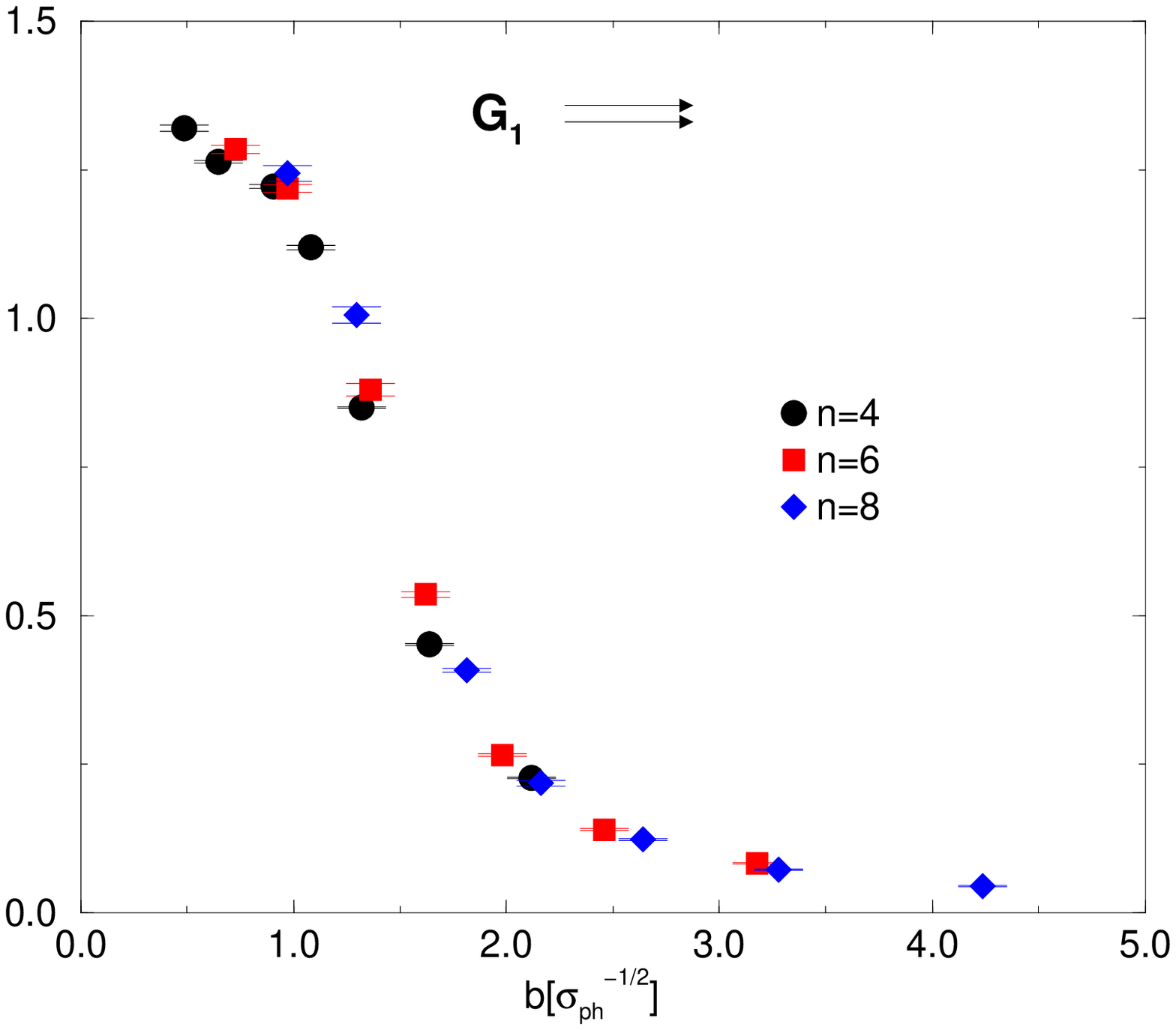}
\end{flushleft}

\vspace{-57mm}

\begin{flushright}
  \epsfxsize=50mm
  \epsfysize=50mm
  \leavevmode
  \epsfbox{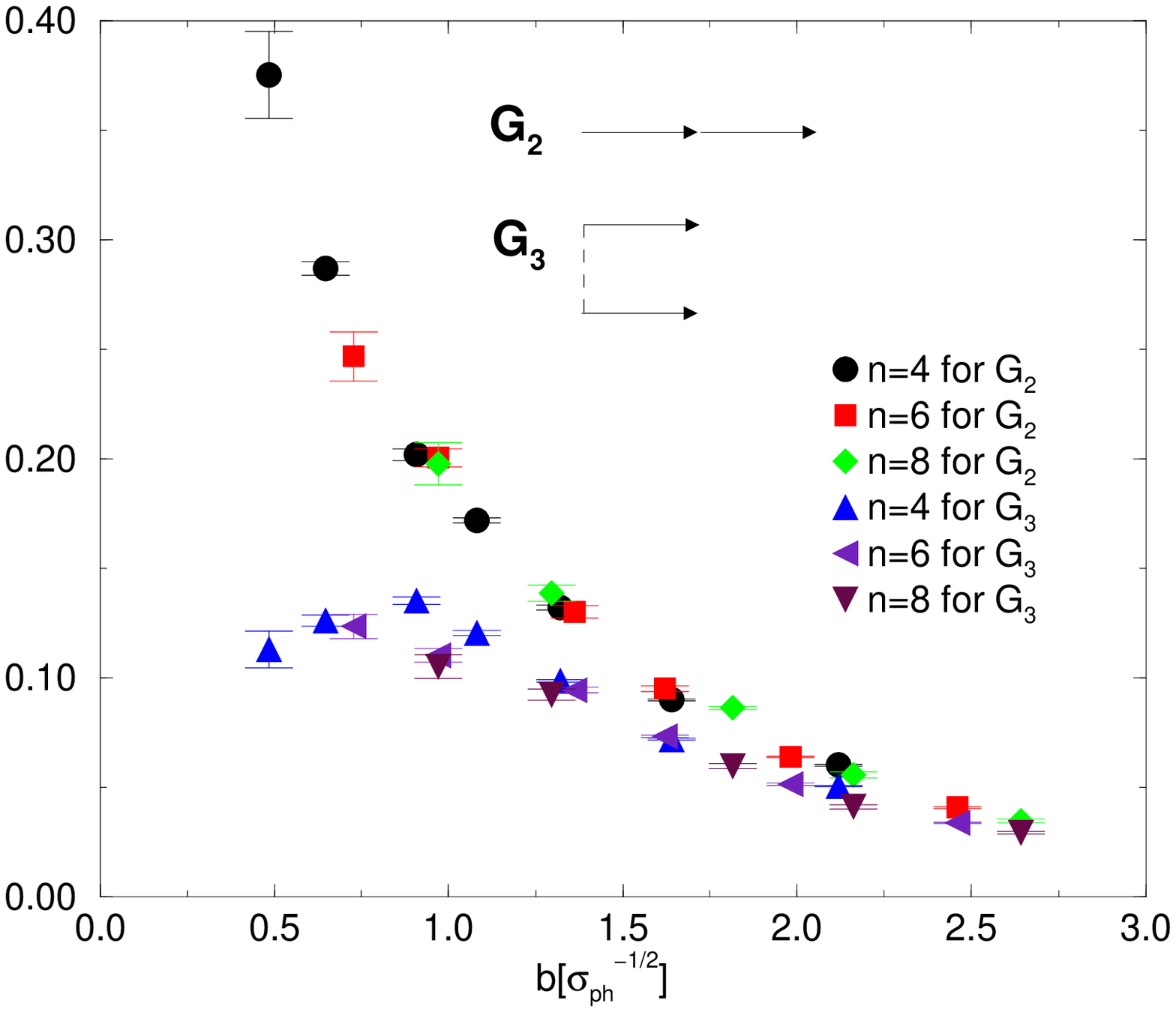}
\end{flushright}

\vspace{.3cm}

Figure 2: The dominant couplings of quadratic interactions 
versus physical length $b$ in SU(3).

\end{minipage}

\item[4.]
The $SU(3)$ case has three types of monopole currents 
$\{k_{\mu}^{(a)}(s), a=1\sim3\}$ with one constraint 
$\sum_a k_{\mu}^{(a)}(s)=0$.
But the behaviors of the effective action are similar to those of the $SU(2)$
case especially for large $b$ region.
\end{itemize}

\section{A perfect operator for a physical quantities}
\label{PfecOpe}

In QCD, the string tension from the 
static potential is 
an important physical quantity. 
A naive abelian Wilson loop 
operator on the coarse lattice is not good \cite{kato98},
because the cut-off effect is of order of the 
lattice spacing of the coarse lattice. 
We should use an improved operator  
on the coarse lattice in order to get the correct values of the physical 
observables. An operator giving a cut-off independent value on RT
 is called {\bf perfect operator}.

\subsection{The method}
The abelian monopole action ${\cal S}[k]$
which is obtained numerically is well approximated by 
quadratic interactions alone. 
We can perform the analytic block spin
transformation along the flow projected on the quadratic coupling
constant plane.
When we define an operator on the fine $a$ lattice, 
we can find a perfect operator along the projected flow 
in the $a\to 0$ limit for fixed $b$.
We adopt the perfect operator on the projected space 
as an approximation of the correct
operator for the action ${\cal S}[k]$ on the coarse $b$ lattice.

\subsection{The operator for the abelian static potential }
First let us 
consider the following abelian gauge theory of the generalized Villain
form on a fine lattice with a very small lattice distance:
\beqn
{\cal S}[\theta,n] = \frac{1}{4\pi^2}\sum_{s,s';\mu>\nu}
(\partial_{[\mu}\theta_{\nu]}(s)+2\pi n_{\mu\nu}(s))
(\Delta D_0)(s-s')
(\partial_{[\mu}\theta_{\nu]}(s')+2\pi n_{\mu\nu}(s')),\nonumber
\eeqn
where $\theta_{\mu}(s)$ is a compact abelian gauge field
and integer-valued tensor $n_{\mu\nu}(s)$ stands for Dirac string.
Both of variables are defined on the original lattice.
Since we are considering a
fine lattice near to the continuum limit, we assume the direction 
symmetry of $D_0$.
In this model, it is natural to use an abelian Wilson loop 
$W({\cal C})=\exp i\sum_{\cal C}(\theta_{\mu}(s),J_{\mu}(s))$
for particles with fundamental abelian charge, where $J_{\mu}(s)$ is
abelian integer-charged electric current.
The expectation value of $W({\cal C})$ is written as 
\begin{eqnarray}
\left< W(C) \right> &=&
  \left<
    \exp\left\{
      i \sum_{s,\mu} J_{\mu}(s)\theta_{\mu}(s)
    \right\}
  \right>
= Z[J]/Z[0], \\
Z[J] &\equiv&
  \int_{-\pi}^{\pi}\prod_{s;\mu}d\theta_{\mu}(s)
  \sum_{n_{\mu\nu}(s)=-\infty}^{+\infty}
  \exp\left\{
    -{\cal S}[\theta,n] + i \sum_{s,\mu}J_{\mu}(s)\theta_{\mu}(s)
  \right\},
\label{ap1.1}
\end{eqnarray}

When use is made 
of BKT transformation\cite{BKT},  the area law term
from the partition function
 (\ref{ap1.1}) 
is equivalent to that from the following monopole expression
\begin{eqnarray}
 \langle W_m({\cal C}) \rangle
&=&\frac{1}{Z}\sum_{k_{\mu}(s)=-\infty
         \atop{\partial^{\prime}_{\mu}k_{\mu}(s)=0}}^{\infty}
  \exp
  \mbra{-\sum_{s,s',\mu} k_{\mu}(s)D_0(s-s')k_{\mu}(s')
       +2\pi i\sum_{s,\mu} N_{\mu}(s)k_{\mu}(s)}, \nonumber\\
N_{\mu}(s,S_J)&=&\sum_{s'}\Delta_L^{-1}(s-s')\frac{1}{2}
\epsilon_{\mu\alpha\beta\gamma}\partial_{\alpha}
S^J_{\beta\gamma}(s'+\hat{\mu}),
\label{pfac:3}
\end{eqnarray}
where $S^J_{\beta\gamma}(s'+\hat{\mu})$ is a plaquette variable satisfying 
$\partial'_{\beta}S^J_{\beta\gamma}(s)=J_{\gamma}(s)$.

\subsection{The perfect operator  
for the static potential on the coarse lattice}
We perform a block spin transformation (\ref{pfac:2}) 
of the monopole currents. 
Let us start from (\ref{pfac:3}). 
The cutoff effect of the operator (\ref{pfac:3}) is $O(a)$ by definition.
The $\delta$-function renormalization group transformation
(\ref{pfac:2})
 can be
done analytically. Taking the continuum limit $a\to 0$, $n\to \infty$
(with $b=na$ is fixed) finally, we obtain the expectation
value of the operator on the coarse lattice with spacing $b=n a$: 
\begin{eqnarray}
\langle W({\cal C}) \rangle
&=&\frac{1}{Z}\langle W_m({\cal C}) \rangle_{cl} 
  \exp\Biggl\{\pi^2 \sum_{s^{(n)},s^{(n)'}\atop{\mu,\nu}}
    \!\!\!\!
    B_{\mu}(s^{(n)})
      D_{\mu\nu}(s^{(n)}-s^{(n)'})
    B_{\nu}(s^{(n)'})
  \Biggr\}
\nonumber \\
\times \!\!\!\!
\sum_{\partial'_\mu K_\mu=0}
\!\!\!\!\!\!
&&
  \exp\Biggl\{
    - S[K_{\mu}(s^{(n)})]
    +2 \pi i \sum B_{\mu}(s^{(n)})D_{\mu\nu}(s^{(n)}-s^{(n)'})
K_{\nu}(s^{(n)'})
  \Biggr\},\ \ \ 
\label{opwil:1}
\end{eqnarray}
where
$B_\mu(s^{(n)}) $ is the renormalized electric source term.
$S[K_{\mu}(s^{(n)})]$ denotes the effective action defined on the coarse
lattice:
\begin{eqnarray}
  S[K_{\mu}(s^{(n)})] =
  \sum_{s^{(n)},s^{(n)'}}\sum_{\mu,\nu}K_{\mu}(s^{(n)})
  D_{\mu\nu}(s^{(n)}-s^{(n)'})K_{\nu}(s^{(n)'}).
\label{pfac:5}
\end{eqnarray}
Here 
 $\langle W_m({\cal C}) \rangle_{cl}$ is defined by
\begin{eqnarray}
\langle W_m({\cal C}) \rangle_{cl}
&=&
  \exp
  \Bigg\{
    -\pi^2 \int_{-\infty}^{\infty}\!\!\!\!d^4xd^4y
    \sum_{\mu}N_{\mu}(x)D_0^{-1}(x-y)N_{\mu}(y)
  \Bigg\}.
\label{opwil:5}
\end{eqnarray}
Since we take the continuum limit analytically, the operator (\ref{opwil:1})
does not have no cutoff effect. For details, see Ref.\cite{fuji00a}.

Performing BKT transformation  on the coarse
lattice, we can get the loop operator for the static potential 
in the framework of the string model:
\begin{eqnarray}
\langle W_m({\cal C}) \rangle
&=& \sum_{\sigma_{\mu\nu}(s)=-\infty
        \atop{\partial_{[\alpha}\sigma_{\mu\nu]}(s)=0}}^{\infty}
  \exp
  \Bigg\{
    -\pi^2\sum_{ s,s'\atop{ \mu\neq\alpha\atop{ \nu\neq\beta } } }
    \sigma_{\mu\alpha}(s)\partial_{\alpha}\partial_{\beta}'
    D_{\mu\nu}^{-1}\Delta_L^{-2}(s-s')\sigma_{\nu\beta}(s')\nonumber\\
&&-2\pi^2\sum_{s,s'\atop{\mu,\nu}}\sigma_{\mu\nu}(s)\partial_{\mu}
    \Delta_L^{-1}(s-s')B_{\nu}(s') 
  \Bigg\}\times \langle W_m({\cal C}) \rangle_{cl}.
\label{opwil:4}
\end{eqnarray}
\subsection{Parameter fitting}
\label{ParaFit}
In order to compare our analysis with the inverse Monte-Carlo method,
we expand 
$D_0(s-s')$ in the monopole action (\ref{pfac:3}) as
$\alpha\delta_{s,s'}+\beta\Delta_L^{-1}(s-s')+\gamma\Delta_L(s-s')$,
where $\alpha$, $\beta$ and $\gamma$ are free parameters.
Then by matching the set of numerically
obtained coupling constants of the monopole action $\mbra{G_i(b)}$ with 
$D_{\mu\nu}(s^{(n)}-s^{(n)'})$ we fix
the free parameters as shown in Ref.\cite{maxim00a}.

\section{Analytic evaluation of physical quantities}
\subsection{The string tension}
Let us evaluate the string tension using the perfect operator 
(\ref{opwil:4}).
The plaquette variable $S_{\alpha\beta}$ in Eq.(\ref{pfac:3}) for the 
static potential $V(Ib,0,0)$ is expressed by
\begin{eqnarray}
S_{\alpha\beta}(z)
&=&
  \delta_{\alpha 1}\delta_{\beta 4}\delta(z_{2})\delta(z_{3})
  \theta(z_{1})\theta(Ib-z_{1})
  \theta(z_{4})\theta(Tb-z_{4}).
\end{eqnarray}
The string
model is in the strong coupling region for large $b$. 
Therefore, we evaluate Eq.(\ref{opwil:4}) by the strong coupling expansion.
As shown in Ref.\cite{maxim00a}, 
the (classical) string tension coming 
from Eq.~(\ref{opwil:5}) is dominant and it becomes in SU(2)
\begin{eqnarray}
\sigma_{cl}=\frac{\pi\kappa}{2} \ln\frac{m_1}{m_2}.
\label{sigma_cl}
\end{eqnarray} 
Using the optimal values $\kappa$, 
$m_1$ and $m_2$, we get 
the string tension $\sqrt{\sigma_{cl}/\sigma_{phys}}$ 
as shown in Table~\ref{stringtension}.
\begin{table}
\begin{center}
\caption{The calculated atring tension}
\vspace{.3cm}
\begin{tabular}{|clll|clll|}\hline
$SU(2)$&&&&$ SU(3)$&&&\\ \hline
$b$       &2.1  &2.9  &3.8 
&$b$ & 2.64&3.27&4.23 \\ \hline
$\sqrt{\frac{\sigma_{cl}}{\sigma_{phys}}}$
          &1.64 &1.56 &1.45
 &$\sqrt{\frac{\sigma_{cl}}{\sigma_{phys}}}$
&1.15&1.22&1.21 \\ \hline
\end{tabular}
\label{stringtension}
\end{center}
\end{table}

\subsection{On the continuum rotational invariance}
\label{ocri}

We here comment on the continuum rotational invariance of the 
quark-antiquark static potential.
We get the static potential $V(Ib,Ib,0)$ in SU(2) can be written as
\begin{eqnarray}
V(Ib,Ib,0) &=& \frac{\sqrt{2}\pi\kappa Ib}{2} \ln\frac{m_1}{m_2}. 
\end{eqnarray}
The potentials from the classical part take only the linear form and the 
rotational invariance is recovered completely even for the nearest $I=1$
sites. 

\section{Gauge independence}
If the color confinement mechanism is due to the condensation 
of the monopoles, it should be gauge independent, since 
the monopole condensation in some special gauge means only abelian charge 
confinement which is quite different from color confinement.
 
Recently Ogilvie has developped a character as well as a 
strong coupling expansions for Abelian
projection and found that gauge fixing is unnecessary, i.e.,
Abelian projection yields string tensions
of the underlying non-Abelian theory even {\it without} gauge
fixing\cite{ogilvie,domi}. Hence at least gauge independence 
of abelian dominance seems to be realized. In the following, we 
show numerical analyses of gauge dependence for a general class of gauge 
between MA gauge and no-gauge fixing. Then we next prove if abelian dominance 
is gauge independent, gauge independence of monopole dominance is
derived.

\subsection{Numerical analyses in a general gauge}
We consider the
Langevin equation with stochastic gauge fixing term\cite{zwanzger,mizu}:
\[
U_\mu(x,\tau+\Delta\tau)=
\omega^{\dagger}(x,\tau){\rm exp}(if^a_\mu t^a)
U_\mu(x,\tau)\omega(x+\hat{\mu},\tau), 
\]
\vspace{-.5cm}
\[
f_\mu^a=-\frac{\partial S}{\partial
A^a_\mu}\Delta\tau+\eta^a_\mu(x,\tau)
\sqrt{\Delta\tau},\nonumber 
\]
\[
\omega(x,\tau)={\rm exp}(i\beta\Delta^a(x,\tau) 
t^a \Delta\tau/2N_c\alpha).\nonumber 
\]
We set 
$ \Delta(x,\tau)=i[\sigma_3,X(x,\tau)] $,
where $X$ is the operator to be diagonalized in MA gauge.
$\alpha=0$ ($\alpha=\infty$) corresponds to the MA gauge fixing
(no gauge fixing).

\setcounter{figure}{2}

\begin{figure}{tbh}
\begin{center}
\caption{Gauge dependence of the string tension}
\vspace{-1.5cm}
\epsfxsize=6cm
\epsfysize=6cm
\leavevmode
\epsfbox{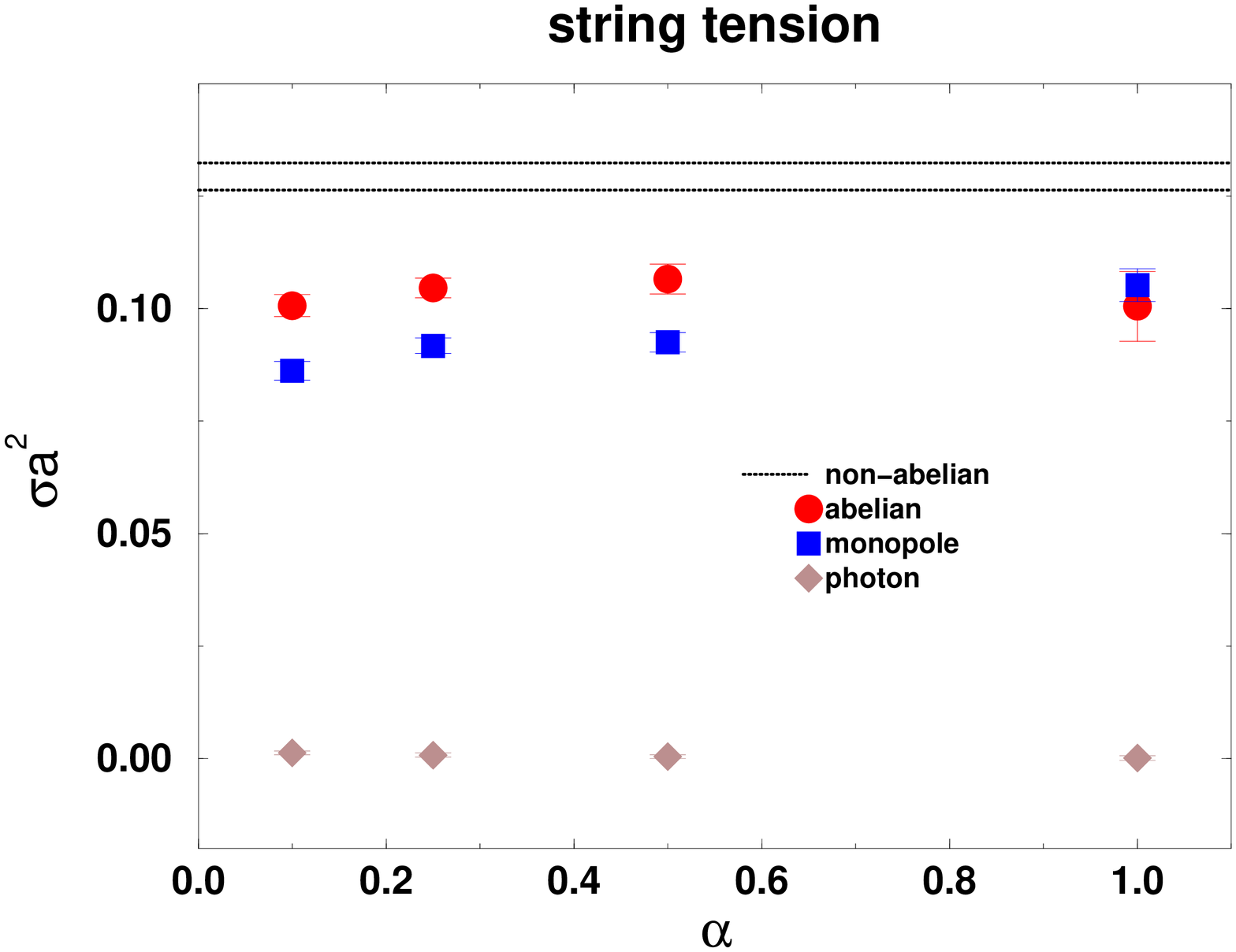}
\label{str_all}
\end{center}
\end{figure}

We have performed 
numerical simulations on $8^3\times 12$ and $16^3\times 24$ lattices
with improved Iwasaki action. See Ref \cite{shoji99} for details.
We get the results shown in Fig.\ref{str_all}. 
Abelian amd monopole dominances are seen for a wide range of $\alpha$.

\subsection{Gauge independence of monopole dominance}
Assume gauge independence of abelian dominance is realized. 
Then we expect the existence of an abelian effective action.
We express the abelian action in terms of the Villain form
$Z=\int_{-\pi}^{\pi} D \theta 
\sum_{n\in {\bf Z}}{\rm e}^{-F[d\theta + 2\pi n]}.$
The general Villain action can be expressed as follows:
\begin{eqnarray*}
&&\hspace*{-.5cm}
{\rm e}^{-(d\theta + 2\pi n,D(d\theta + 2\pi n))
          - F^{'}[d\theta + 2\pi n]} \\
&&\hspace*{-.5cm}=e^{-F^{'}[-i\delta/\delta B]}
{\rm e}^{-(d\theta + 2\pi n ,D(d\theta + 2\pi n))
          + i(B,d\theta + 2\pi n)}\bigg|_{B=0},
\end{eqnarray*}
where $[D,d]=[D,\delta]=0$ are satisfied in the large $\beta$ scaling region.

The abelian Wilson loop ${\rm e}^{i(\theta ,J)}$ is estimated 
with this action, where $J$ is the electric current.
When we use the BKT transformation\cite{BKT}, we get
an action in terms of monopole currents:
\begin{eqnarray*}
Z(J) &=& e^{-F^{'}[-i\delta/\delta B]}\sum_{k\in {\bf Z}, dk=0}
{\rm e}^{-\frac{1}{4}( \delta B , (\Delta D )^{-1}
  \delta B )} \\
&\times&{\rm e}^{\frac{1}{2}( i B , 4\pi \delta \Delta^{-1}k
+ i  (\Delta D )^{-1} dJ )} 
{\rm e}^{-4\pi^{2} ( k , \Delta^{-1}D  k )}
{\rm e}^{2\pi i ( \delta \Delta^{-1}k , S )}
{\rm e}^{- \frac{1}{4}( J , ( \Delta D )^{-1} J )}
\bigg|_{B=0} 
\end{eqnarray*}

1)Electric-electric current $J-J$ interactions (with no monopole $k$) 
come from the exchange
of regular photons and have no line singularity leading to a linear 
potential. Hence
the linear potential of abelian Wilson loops is due to 
the monopole contribution alone. 
Monopole dominance is proved 
from abelian dominance.
\noindent
2)The linear potential comes only from  
$\exp(2\pi i (\delta \Delta^{-1}k , S))$. 
The surface independence of the 
static potential is assured due to the 4-d linking number.

The author is grateful 
to S.Kato, S.Fujimoto, M. Chernodub and M.Polikarpov 
for useful discussions and collaboration.
This work is supported 
by the Supercomputer Project (No.98-33, No.99-47)
of High Energy Accelerator Research Organization (KEK)
and the Supercomputer Project 
of the Institute of 
Physical and Chemical Research (RIKEN).
The author acknowledges also the financial support from  
JSPS Grant-in Aid for Scientific Research (B) (No.10440073 and
No.11695029).


\begin{thebibliography}{99}
\bibitem{fuji00a} S.Fujimoto, S.Kato and T.Suzuki,  {\it Phys.~Lett.} {\bf
B476} (2000) 437.
\bibitem{maxim00a} M.Chernodub {\it et al.}, to appear in Phys. Rev. D.
\bibitem{yama00a} K.~Yamagishi {\it et al.}, to appear in JHP.
\bibitem{shoji99} F.~Shoji {\it et al.}, {\it Phys.~Lett.} {\bf
B476} (2000) 199.
\bibitem{lat99} T.~Suzuki, to appear in {\it Nucl.~Phys.}(Proc.~Suppl.). 
\bibitem{'thooft} G.'t~Hooft, {\it Nucl.~Phys.} {\bf B190} (1981)
455.
\bibitem{domi} T.~Suzuki and I.~Yotsuyanagi, {\it Phys. Rev.}
{\bf D42} (1990) 4257; {\it Nucl.~Phys.} {\bf B} {\it (Proc.~Suppl.)}
{\bf 20} (1991) 236; S.~Hioki {\it et al.}, {\it Phys.~Lett.} {\bf
B272} (1991) 326 and references therein.
\bibitem{Kronfeld} A.S.~Kronfeld {\it et al.}, {\it Phys.~Lett.} {\bf
198B} (1987) 516; A.S.~Kronfeld, G.~Schierholz and U.J. Wiese, {\it
Nucl.~Phys.} {\bf B293} (1987) 461.
\bibitem{shiba_suzuki} H.~Shiba and T.~Suzuki,
{\it Phys.~Lett.} {\bf B343} (1995) 315,
{\it Phys.~Lett.} {\bf B351} (1995) 519
and references therein.
\bibitem{Swendsen} R.H.~Swendsen, {\it Phys.~Rev. ~Lett.} {\bf 52}
(1984) 1165; {\it Phys.~Rev.} {\bf B30} (1984) 3866,3875.
\bibitem{nakam} S.~Kato, S.~Kitahara, N.~Nakamura and T.~Suzuki,
{\it Nucl.~Phys.} {\bf B520} (1998) 323.
\bibitem{D_T} T.A. DeGrand and D. Toussaint, Phys. Rev. 
{\bf D22} (1980) 2478.
\bibitem{ivanenko} T.L. Ivanenko, A.V. Pochinskii and
M.I. Polikarpov, Phys. Lett. {\bf B 252} (1990) 631.
\bibitem{kato98} S. Fujimoto et al, Nucl.Phys.B(Proc.Suppl)73(1999) 533.
\bibitem{BKT} V.L.~Beresinskii, {\it Sov.~Phys.~JETP} {\bf 32}
(1970) 493.
J.M.~Kosterlitz and D.J.~Thouless, {\it J.~Phys.},{\bf
C6} (1973) 1181.
\bibitem{teper98} M. Teper, hep-th/9812187
\bibitem{ogilvie} M.C. Ogilvie,Phys.Rev. D59 (1999) 074505.
\bibitem{zwanzger} D.Zwanziger,
Nucl. Phys. {\bf B192} (1981) 259.
\bibitem{mizu} M.Mizutani and A.Nakamura,
Vistas in Astronomy, {\bf 37}, 305, 1993, Pergamon Press;
Nucl.Phys. B(Proc.Suppl.)34, 253, 1994.


\end{thebibliography}
\end{document}